\documentclass[12pt]{article}

\usepackage{psfig}
\usepackage{multicol}
\usepackage[dvips]{graphicx}
\usepackage{amssymb}
\usepackage{indentfirst}
\usepackage{latexsym}

\title{\bf Proximity effect in planar TiN-Silicon junctions}

\author{D. Quirion\footnote{Corresponding author, email : quirion@drfmc.ceng.cea.fr}, F. Lefloch and M. Sanquer
\\D\'epartement de recherche fondamentale et de la mati\`ere condens\'ee
\\Service de physique statistique, magn\'etisme et supraconductivit\'e
\\CEA/Grenoble - 17 Avenue des Martyrs
\\38054 Grenoble cedex 9, France.}

\begin{document}

\maketitle

\begin{abstract}

We measured the low temperature subgap resistance of titanium nitride (superconductor, T$_{c}$=4.6K)/highly doped silicon (degenerated semiconductor) SIN junctions, where I stands for the Schottky barrier. At low energies, the subgap conductance is enhanced due to coherent backscattering of the electrons towards the interface by disorder in the silicon (''reflectionless tunneling''). This Zero Bias Anomaly (ZBA) is destroyed by the temperature or the magnetic field above $250 \,mK$ or $0.04 \,T$ respectively. The overall differential resistance behavior (vs temperature and voltage) is compared to existing theories and values for the depairing rate and the barrier transmittance are extracted. Such an analysis leads us to introduce an effective temperature for the electrons and to discuss heat dissipation through the SIN interface.

\end{abstract}

\section{Introduction}

When a diffusive normal metal (or doped semi-conductor) (N) is electrically connected to a superconductor (S), electron-electron correlations (i.e. a finite pair amplitude) are induced in the normal metal through Andreev reflections \cite{Andreev64}: this is known as the proximity effect. The subgap conductance is a sensitive measurement of these correlations and a lot of recent works has been devoted to this subject, especially with the development of nanotechnologies. In the original version of the proximity effect \cite{Deutscher69}, the strength of the induced correlations depends on the barrier transmittance $\Gamma$ at the interface and their extension is limited either by the thermal length $L_{T}=\sqrt{\frac{\hbar D}{2\pi k_{B}T}}$ or by $L_{V}=\sqrt{\frac{\hbar D}{eV}}$ in a non-equilibrium situation (D is the diffusion constant of the normal metal); the phase breaking length $L_{\varphi}$ or the geometry of the sample at small distances were not taken into account.
 
Emergence of mesoscopic physics has enlightened the role of these effects, both when the contact between the superconductor and the metal is very good (high transmittance $\Gamma \approx 1$) and in the opposite case (tunnel junctions). In high transmissive contacts, when $L_{T}$ becomes larger than the typical sample length $L$ (supposed smaller than $L_{\varphi}$), the re-entrance phenomenon takes place \cite{Courtois99a}, which cancels out the standart proximity effect: instead of a steadily increase of the subgap conductance as T decreases,  the conductance has a maximum at $T_{0}$ such that $L_{T_{0}} \simeq L$ and  decreases at lower temperature, theoretically recovering its normal value for $T=0K$ (without interaction in the normal metal) \cite{Artemenko79}. On the other hand, when the barrier is rather strong (low transmittance $\Gamma \ll 1$), the low temperature subgap conductance is small since the dominating Andreev reflection is a two-particle process which scales as $\Gamma^{2}$. The subgap conductance for any value of $0 \le \Gamma \le 1$ is described by the BTK theory \cite{BTK}, without taking into account coherence effects (ballistic normal metal). Coherent effects are also observed experimentally for $\Gamma \ll 1$: when the  Andreev reflected hole that traces back the incoming electron path  is coherently backscattered towards the interface, the amplitudes for successive Andreev reflections add constructively. The subgap conductance is greatly enhanced as if the retroreflected hole did not feel the barrier  (``reflectionless tunneling'') \cite{vanWees92}. The conductance shows a peak at low energies, whose amplitude depends on the balance between the tunnel transparency and the rate of coherent backscattering. For a diffusive normal metal, the peak in the conductance increases when the  coherent resistance increases and is maximum when the coherent normal resistance equals the barrier resistance. Eventually, if the coherent normal resistance increases above the barrier resistance, the situation evolves generically to the first case (good NS contact) and the conductance shows a peak at finite energy \cite{Poirier97}.

So far, reflectionless tunneling has essentially been observed in Superconductor/Semiconductor junctions (S/Sm) (the only exception is the NS-SQUID devices by Pothier et al. \cite{Pothier94}). In these systems, the low transparency of the  interface is due to a Schottky barrier and to the mismatch of the Fermi velocities. The transparency of these junctions is intermediate  (typically $\Gamma_{S/Sm} \simeq 10^{-3}$) between a very good interface (transmission coefficient $\Gamma \simeq$1) and an oxide barrier ($\Gamma \simeq 10^{-6}$). In semiconductors, annealing and surface cleaning processes or even the absorption of dopands by the electrode  produce an increase of the sheet resistance of the semiconductor  below the interface overlap. As reflectionless tunneling is a balance between the electron probabilities of crossing the barrier and backscattering to the interface, S/Sm systems are subject  to  this effect. But the superconducting material can also be weakened near the interface and consequently the BCS density of states of the superconductor can be smoothed by pair-breaking processes or creation of states below the gap. Zero-bias anomalies are always associated to such smooth conductance-voltage characteristics \cite{Poirier97,Kastalsky91,Magnee94,Bakker94}. Because this effect is very sensitive to the microscopic parameters near the interface as well as to the energy of carriers, one can take advantage of its observation both to feature the S/Sm contact and to investigate the thermalization of electrons in the normal part. The latter point has concentrated a lot of recent works in out-of-equilibrium mesoscopic normal conductors  \cite{Pothier97,Baselmans99,Henny97,Steinbach96}. The problem of carriers thermalization  is even more crucial in S/N or S/N/S junctions because of the Andreev thermal resistance at the interface \cite{Jehl99,Jehl00,Hoss99}. It is also of practical importance for superconducting bolometers or Josephson field effect transistors. 
 
In this article, we report the observation of zero-bias anomalies in titanium nitride/heavily doped silicon junctions at very low temperature (down to 30mK). In a first part, the samples are characterized with various measurements (contact resistance, sheet resistance, weak localization). In a second part, the differential resistance of the junctions is measured as a function of temperature, voltage and magnetic field. The third part is devoted to the quantitative comparison between the observed zero bias anomaly and the theory for a planar SIN junction. A good agreement is obtained for temperature behavior, allowing us to extract the depairing rate and the barrier transmittance of the junctions. However, discussion on the voltage response of the junctions leads us to consider the effective temperature of the carriers in the silicon, which is well above the phonon bath temperature. This overheating effect is discussed in the fourth part. 

\section{Samples and materials characterisation}
\label{materials}

\begin{figure}
\begin{center}
\psfig{figure=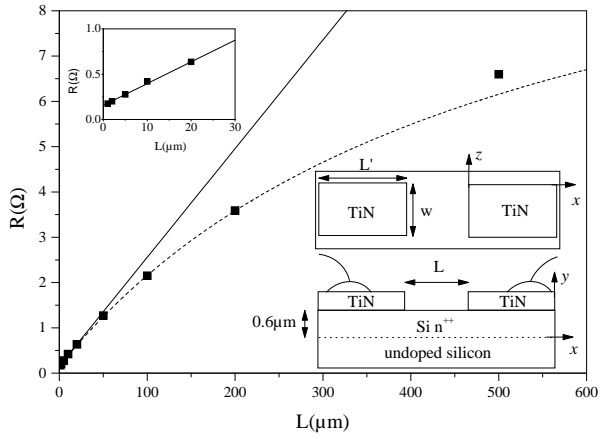,width=100mm}
\caption{Tranverse Length Method (TLM) measurements at 4K for different lengths $L$. In upper inset, the same curve for small $L$. We deduce the sheet resistance of the bulk silicon $R_{\Box}=24\Omega$ and the normal state resistance of an interface $R_{NN}=7.8\ 10^{-2} \Omega$ for a w=1000$\mu$m width junction. The dashed curve is obtained by adding two squares of silicon in parallel to take into account the conduction through the silicon layer outside the gap between the TiN electrodes. In bottom inset, schematic view of the samples.}
\label{fig:tlm}
\end{center}
\end{figure}

The samples are fabricated from a TiN(1000\AA)/Si $n^{++}$ (P doped on $d_{Si:P}$=0.6$\mu$m) bilayer deposited on a $Si \, 8'$ wafer. First, the wafer is oxidized on 13nm. Phosphorus is then implanted at 15 keV, $2.10^{15} cm^{-2}$, followed by a recristallisation heat treatment at 650$^{\circ}$C and a $30$ minutes activation/diffusion treatment at 1050$^{\circ}$C with oxygen. The substrate is deoxidized and a 10nm Ti and 100nm TiN bilayer is deposited. Optical lithography defines a Tranverse Length Method (TLM) pattern with various distances $L=1, 2, 5, 10, 20, 50, 100, 200$ and $500\mu m$ between large TiN pads (typically 1000$\times$1000$\mu m^{2}$) (see bottom insert of figure \ref{fig:tlm}). The TiN/Ti is etched and a final heat treatment at 720$^{\circ}$C in N$_{2}$ atmosphere during 20 seconds provides TiN densification and forms a 40nm thick TiSi$_{2}$ layer. Nevertheless, the zero bias anomaly reported in this work appears rather insensitive to this heat treatment. On some samples, another etching is performed to define two TiN pads fingers with lateral dimensions w=10$\mu$m facing each other and connected to larger TiN reservoirs. 

Although titanium nitride has been used for many years in microelectronics as a diffusion barrier, ohmic contact and gate electrode in field effect transistors \cite{Wittmer82}, its  superconducting properties have been thoroughly studied only recently \cite{Lefloch99}. Its transition temperature is T$_{c}$=4.6 K for our samples. STM measurements give a gap of 250$\mu$V at T=1.4K \cite{Chapelier99}. The relation between the  superconducting gap and the transition temperature departs from the BCS theory probably because titanium nitride is a granular superconductor. Its room temperature resistivity is 85$\mu\Omega$.cm.

Doped Si:P has been studied for a long time because of its great importance in microelectronics. Alexander et al. \cite{Alexander68} evaluated the Mott-transition donor concentration $n_{c}=3.10^{18} cm^{-3}$ and the concentration at which the Fermi level of the electron system passes into the conduction band of the host crystal $n_{cb}=2.10^{19} cm^{-3}$. Heslinga et al. \cite{Heslinga92} studied the inelastic lifetime in heavily doped Si:P. They found $1/\tau_{in}(s)=1.1\ 10^{9} \ \ T(K)^{2.2}$ at $n=2.10^{19} cm^{-3}$ from T=1.2 to 4K. In our samples, doped silicon Si:P forms the normal part of the junction with a donor concentration  $n_{e}$=2.10$^{19}$ cm$^{-3}$ over a depth of P implantation  $d_{Si:P}$=0.6$\mu$m. 
From the TLM geometry, we estimate the sheet resistance of the doped silicon and the resistance $R_{NN}$ of the interfaces at T=4K:
\begin{equation}
R=\frac{R_{\Box}}{N_{\Box}} + 2R_{NN} 
\label{eq:tlm}
\end{equation}
where $R_{\Box}$ is the sheet resistance of the doped silicon, $N_{\Box}=w/L$ the number of squares of silicon in parallel between the two SIN contacts, L' and w the length and width of the TiN overlap and L the distance between the two TiN electrodes (see inset figure \ref{fig:tlm}).  Following Giaever's calculation \cite{Giaever69}, $R_{NN}=\sqrt{R_{b} R_{\Box}^{*}}/w$ is the normal state resistance per contact (determined at temperatures just below $T_{c}$, to eliminate the resistance of the TiN pads), with $R_{b}$ the barrier resistance expressed in $\Omega .\mu m^{2}$ and $ R_{\Box}^{*}$ the sheet normal resistance below the overlap (eventually $ R_{\Box}^{*} \ge  R_{\Box}$ for the bare film) . From equation \ref{eq:tlm}, we see that the resistance is linear with the distance $L$. This is what we observed at small $L$ on figure \ref{fig:tlm}. For $L > 50\mu m$, dispersion of the current lines on the sides of the normal part of the S/N/S system becomes noticeable (the TiN pads are on top of an infinite silicon layer). Therefore, we can recover the experimental curve by adding in this case two squares of silicon to $N_{\Box}$.  

The linear fit at small lengths (using $w=1000 \mu m$) gives at 4K (see inset figure \ref{fig:tlm}):
\begin{equation}
R (\Omega)=0.024\ L(\mu m) + 0.156
\end{equation}
We deduced a sheet resistance of the bulk silicon (between the interfaces): $R_{\Box}=24\Omega$. The resistivity at 4K is then $\rho$=14.4 $\mu\Omega$.m in very good agreement with reference \cite{Alexander68}. The normal resistance of each interface is $R_{NN}=7.8\ 10^{-2} \Omega$ for $w$=1000$\mu$m.

Given the effective masses of silicon $m^{\ast}=0.321m_{e}$ (with $m_{t}^{\ast}=0.19m_{e}$ and $m_{l}^{\ast}=0.916m_{e}$) and the valley degeneracy N=6, we used a free electrons model to calculate the parameters of the doped bare silicon: 
\begin{eqnarray}
k_{F}&=&\Big(\frac{\pi^{2} \ n_{e}}{2} \Big)^{1/3}=4.62\ 10^{8} m^{-1} \textrm{ and } \lambda_{F}=13.6nm \\
\ell_{e}&=& \frac{\hbar k_{F}}{\rho e^{2} n_{e}}=6.6 nm \textrm{ and } k_{F} \ell_{e}=3 \\
D&=&\frac{1}{3} v_{F} \ell_{e}=3.67\ 10^{-4} m^{2}.s^{-1} \label{eq:D}
\end{eqnarray}
Those values ensure that the doped silicon is in the metallic regime, since $k_{F} \ell_{e}>1$ and $n_{e}>n_{c}$.

\begin{figure}
\begin{center}
\psfig{figure=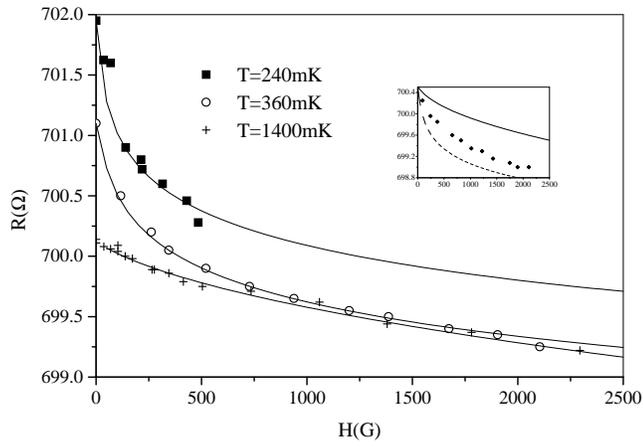,width=100mm}
\caption{Magnetoresistance of a wide wire of Si $n^{++}$ measured for various temperatures. We deduced $L_{\varphi}$=420, 620 and 1000nm respectively at 1400, 360 and 240mK. In inset, the magnetoresistance at 700mK shows the cross-over between 2D and 3D regimes (curves at $L_{\varphi}$=$d_{Si:P}$=600nm depth of doped silicon).}
\label{fig:magnetor}
\end{center}
\end{figure}

Finally, we measured the magnetoresistance of the bulk silicon using a long and wide bar ($20 mm \times 1mm$) of Si $n^{++}$, at various temperatures (figure \ref{fig:magnetor}). We fitted the experimental curves with the 2D and 3D theories of weak localization \cite{Altshuler88}:

\begin{eqnarray}
\sigma&=&\sigma_{Boltzmann}+\Delta \sigma \\
\Delta \sigma&=&\frac{e^{2}}{2\pi^{2}\hbar}f_{2}(2\frac{L_{\varphi}^{2}}{L_{H}^{2}}) \textrm{ at 2D } (L_{\varphi} \ge d_{Si:P})\\
 &=&\frac{e^{2}}{2\pi^{2}\hbar L_{H}}f_{3}(2\frac{L_{\varphi}^{2}}{L_{H}^{2}})  \textrm{ at 3D } (L_{\varphi} \le d_{Si:P})
\end{eqnarray}
with 
\begin{eqnarray*}
f_{2}(x)&=&\ln (x)+\psi (x) \\ f_{3}(x)&=&\sum_{n=0}^{\infty}(2(\sqrt{n+1+\frac{1}{x}}-\sqrt{n+\frac{1}{x}})-\frac{1}{\sqrt{n+\frac{1}{2}+\frac{1}{x}}})
\end{eqnarray*}
$L_{H}=\sqrt{\hbar/(2eH)}$ and $L_{\varphi}$ are the magnetic and the phase-breaking lengths and $d_{Si:P}$ the depth of doped silicon. $\psi (x)$ is the digamma function. We deduced $\tau_{\varphi}=L_{\varphi}^{2}/D$ for various temperatures (see table \ref{tab:tauphi}). As we were not able to fit the experimental curve at T=700mK, we concluded that the cross-over between the 2D and 3D regimes lies in this temperature range (corresponding to $L_{\varphi}\simeq d_{Si:P}$). For lower temperatures, we used the 2D theory and for T=1.4K, the 3D theory. At T=1.4K, we measured $\tau_{\varphi}=0.48ns$, in good agreement with Heslinga et al \cite{Heslinga92} who obtained $\tau_{\varphi}=0.58ns$ at T=1.2K. At lower temperature, these results can be compared to the expression given by Altshuler et al. \cite{Aleiner99}, where only $R_{\Box}$ enters as a parameter:
\begin{equation}
\frac{\hbar}{\tau_{\phi} (2D)}=k_{b}T \frac{e^{2}}{h} R_{\Box} ln(\frac{h}{2e^{2}R_{\Box}})
\label{eq:tauphi}
\end{equation}
which gives $\tau_{\varphi}=1.3ns$ at 1K (see table \ref{tab:tauphi}), in relative good agreement with our experimental results.

\begin{table}[h]
\begin{center}
\begin{tabular}{|c|c|c|}
\hline
Temperature (mK) & $\tau_{\phi}$ (ns) (exp.) & $\tau_{\phi}$ (ns) (calculated eq. \ref{eq:tauphi}) \\
\hline \hline
1400 & 0.48 (3D) & 0.9 \\
700 & cross-over 2D-3D & 1.9 \\
360 & 1 (2D) & 3.6 \\
240 & 2.7 (2D) & 5.4 \\
\hline
\end{tabular}
\caption{Coherence time for various temperatures deduced from figure \ref{fig:magnetor} and reference \cite{Aleiner99}.}
\label{tab:tauphi}
\end{center}
\end{table}

As distances between the TiN electrodes are larger than or of the order of the phase-breaking length for all studied samples, as far as the coherence is concerned, one can divide the samples into three parts: the silicon resistance and the two interfaces. No coherence effect should link both interfaces and they could be studied as two N/S systems in series. This is confirmed in the next section.

\section{Zero bias anomaly in TiN/Si planar junctions}

\begin{figure}
\begin{center}
\psfig{figure=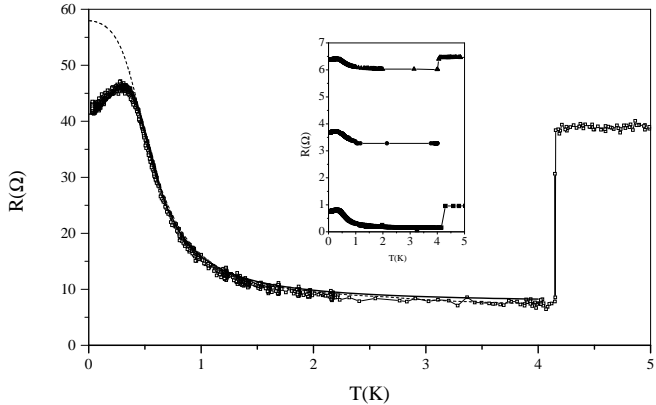,width=100mm}
\caption{ Resistance per contact versus temperature at zero bias for a 20$\times$10 $\mu m^{2}$ sample. The fitted curve used the following parameters. BTK (dashed curve): $\Delta=0.21meV$, $\Gamma_{s}/\Delta=0.11$, Z=5.3 and $R_{NN}=7.45 \Omega$. Volkov (solid curve): $\Delta=0.22meV$, $\Gamma_{s}/\Delta=0.15$, $\gamma/\Delta=0.27$,  $\epsilon_{b}/\Delta=0.009$ and $R_{NN}=7.45 \Omega$. In inset, total resistance-temperature characteristics for samples of various lengths L and $w=1000 \mu m$: from top to bottom, L=1, 200 and 500$\mu$m.}
\label{fig:rt}
\end{center}
\end{figure}

We measured the sample resistance versus temperature (down to $30 mK$) by a standard four probes lock-in technique ($I_{ac}=10$ nA at 180Hz). We studied samples of different lengths (from L=1$\mu$m to 500$\mu$m) and they all show the same behaviour at low temperature (see inset figure \ref{fig:rt}). Just above 4K, we observe a step in the resistance due to the superconducting transition of the TiN electrodes.
Independently, by measurements on a long and wide bar of doped silicon, we have checked the silicon resistance does not depend (or only slightly) on temperature and voltage (see figure \ref{fig:magnetor} and part \ref{materials}). Since we know the normal resistance from TLM measurements, we can plot the resistance per contact: $R_{c}(T,V)=\frac{1}{2}(R_{total}(T,V)-R_{Si})$ where $R_{Si}$ is the resistance of the silicon between the two contacts. So we are able to plot the resistance {\it per contact} $R_{c}$ as a function of the voltage drop {\it at the interface} $V_{i}$. The voltage drop at each interface is $V_{i}=\frac{1}{2}(V_{total}-R_{Si}I)$ where $I$ is the applied DC current.

In the rest of the paper and in figure \ref{fig:rt}, we plot $R_{c}$.
Below T=4.2K, the resistance increases as expected for an SIN junction. Anticipating the discussion in the section 4, its temperature dependence is well fitted within the BTK model \cite{BTK} above $T \simeq 400mK$. The resistance increases only by a factor of 6 between 4K and 400mK, indicating a strong departure from the sharp BCS density of states. This is taken into account via a high damping factor $\Gamma_{S}$ in the superconductor ($\Gamma_{S}/\Delta=0.11$ in the BTK adjustment) while the barrier is rather opaque (transparency $3.4\ 10^{-2}$). At 250 mK, it shows a maximum and then decreases. This can not be explained as a precursor of a Josephson current, since this effect is observed for distances between superconducting electrodes much larger than the coherence length $L_{T}=\sqrt{\hbar D/(2\pi k_{B}T)} = 120 nm$ at 30mK (see figure \ref{fig:rt}). We interpret this decrease of the resistance at low temperature as due to reflectionless tunneling \cite{Kastalsky91,vanWees92}.

\begin{figure}
\begin{center}
\psfig{figure=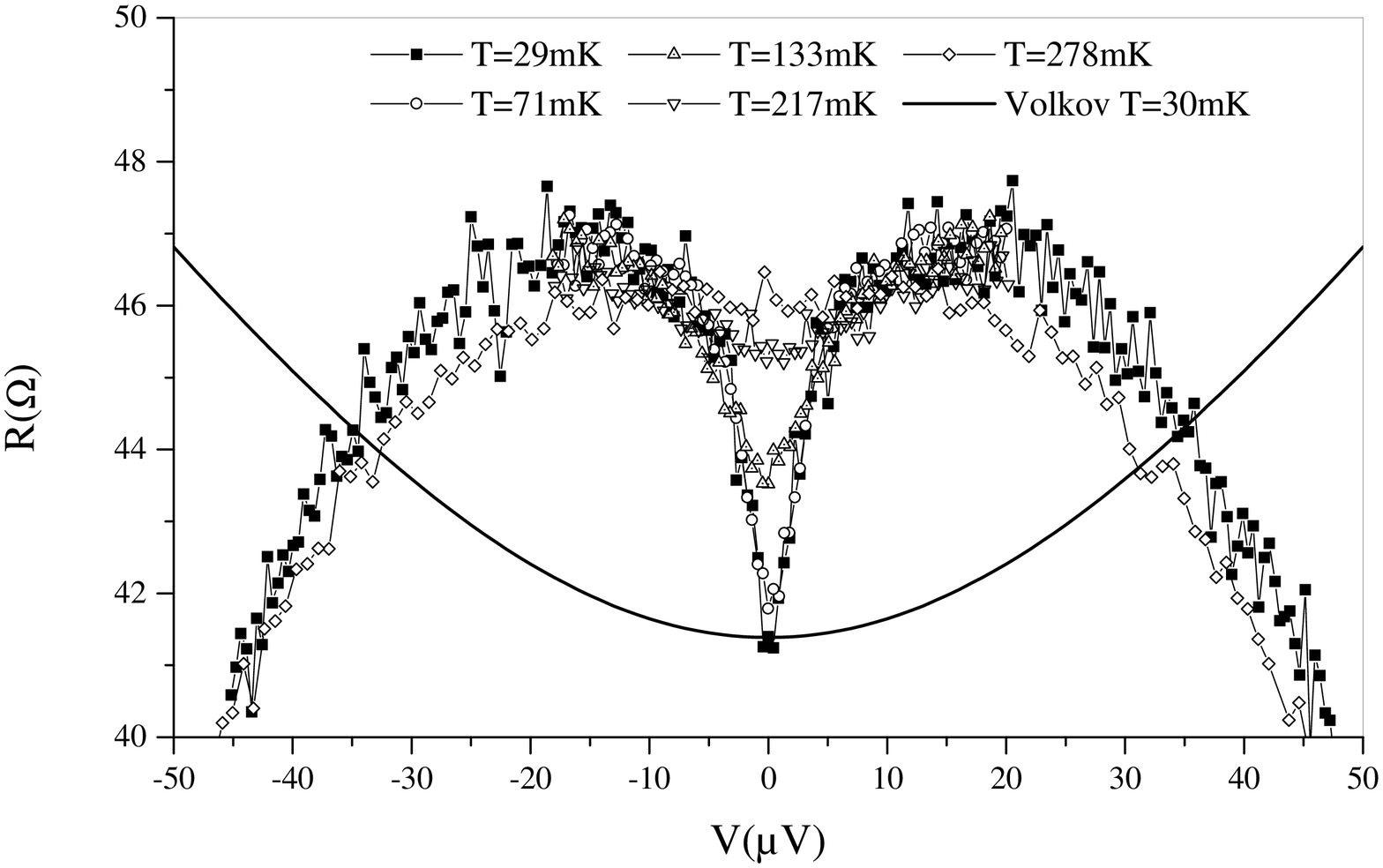,width=100mm}
\psfig{figure=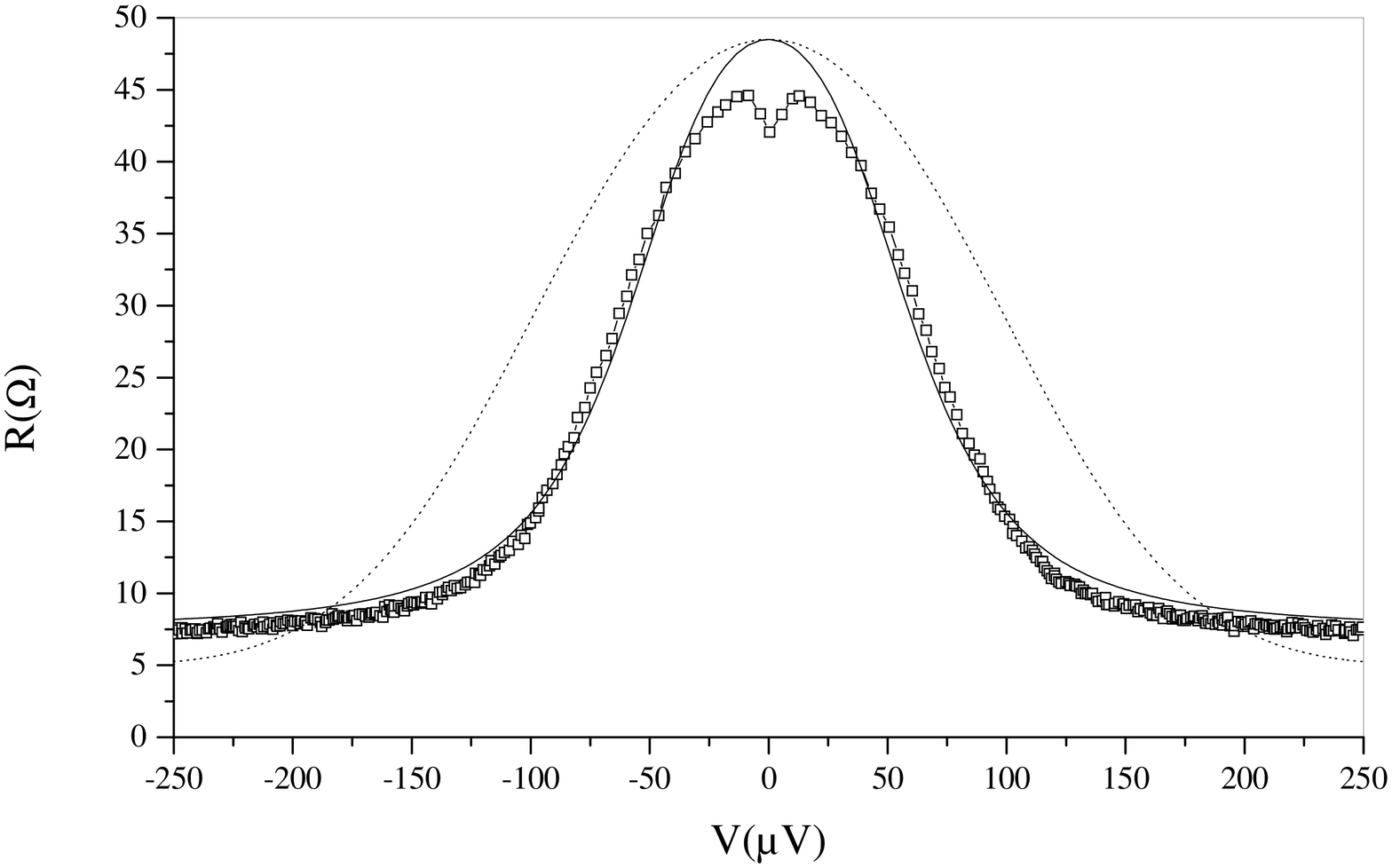,width=100mm}
\caption{TOP: Resistance per contact versus voltage for different temperatures for a 20$\times$10 $\mu m^{2}$ sample and the theoretical curve at 30mK after Volkov model (parameters extracted from the temperature dependence of R(V=0)). BOTTOM: Resistance per contact versus voltage at large voltages, BTK fit with T=320mK (dotted line) and BTK fit with a Wiedemann-Franz law (full line) for a 1$\times$10 $\mu m^{2}$ sample. The BTK parameters are also extracted from the temperature dependence of R(V=0): $\Delta=0.22meV$, $\Gamma_{s}/\Delta$=0.12, Z=5.3, $R_{NN}=7.45\Omega$ and $T_{0}=320mK$.}
\label{fig:r-v}
\end{center}
\end{figure}

We also measured the differential resistance versus  DC voltage at different temperatures (see figure \ref{fig:r-v}) by applying a DC and ac current and by recording both the DC and ac voltages.

Except at low voltages and low temperatures, the $R_{c}(V)$ characteristics show the same smooth behavior for all samples: as the voltage is increased, the resistance steadily decreases and reaches its normal value above the gap voltage: there is no shoulder around $eV = \Delta$. No shoulder is observed either at the gap voltage  for previously studied Sm/S contacts exhibiting reflectionless tunneling \cite{Kastalsky91}. At low temperature, the resistance shows a dip (ZBA) consistent with the reflectionless tunneling regime. When the temperature is increased (see figure \ref{fig:r-v}), the ZBA amplitude vanishes and disappears at 250mK. Moreover, as a function of the voltage, the ZBA disappears on the same energy scale than as a function of temperature, i.e. roughly 20$\mu V$.        

\begin{figure}
\begin{center}
\psfig{figure=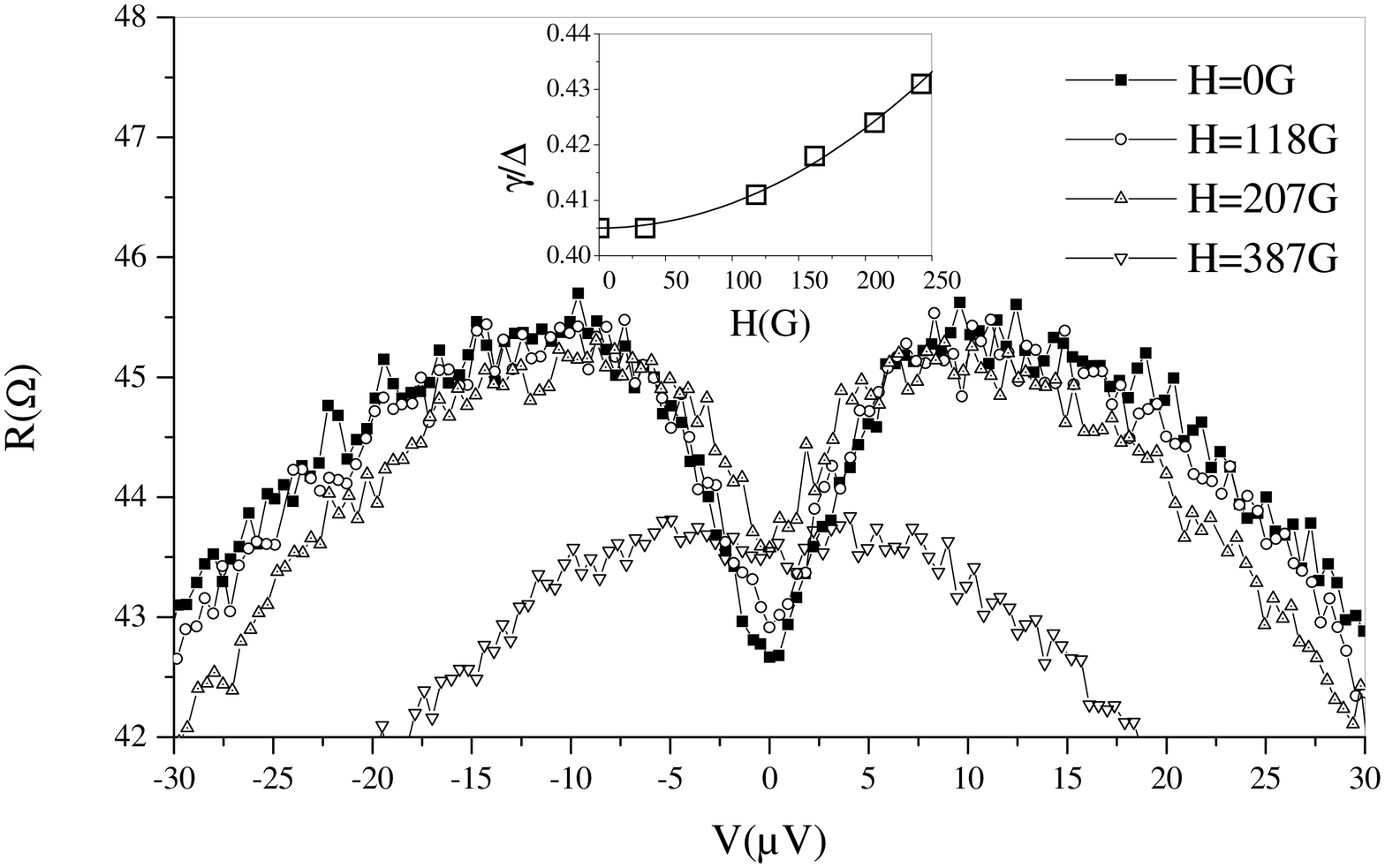,width=100mm}
\psfig{figure=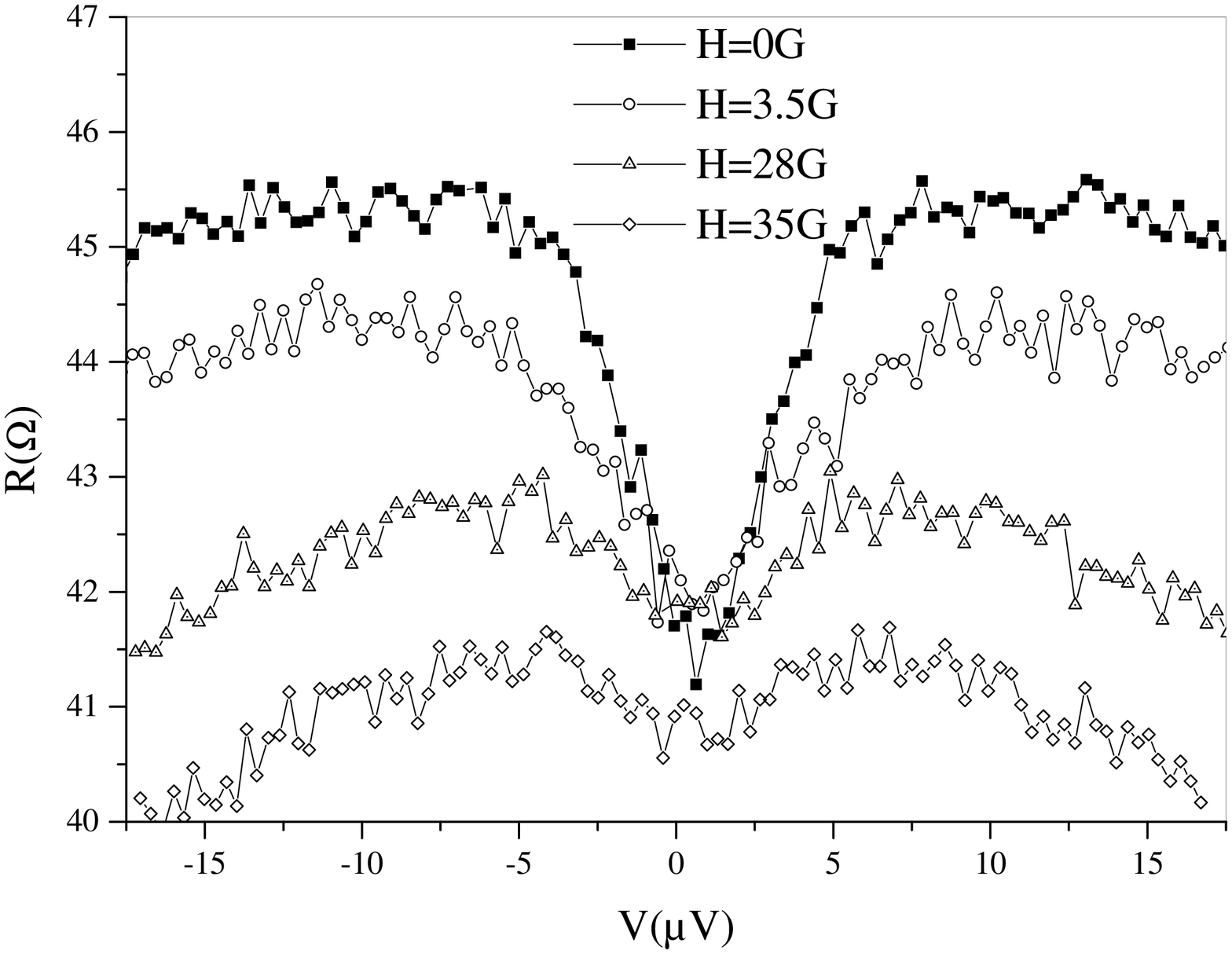,width=100mm}
\caption{Resistance per contact versus voltage for various magnetic fields applied parallel (top figure) and perpendicularly (bottom figure) to the interface for a 1$\times$10 $\mu m^{2}$ sample. In inset of the top figure, the dependence of the pair-breaking energy $\gamma_{in}$ versus the applied magnetic field is obtained from the evolution of R(V=0) versus H. The fit gives the pair breaking parameter $\gamma$ deduced from the evolution of R(V=0) versus H: $\gamma_{in}(H)/\Delta=0.405 + 4.5\ 10^{-7} H(G)^{2}$ \cite{Volkov93}.}
\label{fig:r-v-h}
\end{center}
\end{figure}

Finally, we measured the differential resistance versus the DC voltage for various magnetic fields (see figure \ref{fig:r-v-h}). Two behaviors can be distinguished. When the magnetic field is perpendicular to the S/N interface (along $y$, see figure \ref{fig:tlm}), the resistance decreases at very small field and the ZBA is destroyed at 30G. On the contrary, for weak applied magnetic field parallel to the interface (along $z$), the resistance background is unchanged at small field and only the ZBA diminished. Then, above 200G, the overall resistance decreases. The ZBA is divided by a factor two for $H_{c}^{z} \simeq 200G$ and completely disappears at 400G. 

\section{Proximity effect in planar SIN junctions}

Many theoretical works have been completed since Kastalsky experiments \cite{Beenakker95, Hekking94, Lambert98}.  Volkov \cite{Volkov93} considered planar SIN junctions and, using a theory based on the Usadel equation, he calculated the conductance-voltage characteristics of this system at any voltages and temperatures. Assuming $\epsilon_{b}<<\gamma_{in}$, with $\epsilon_{b}$ the barrier energy (see below) and $\gamma_{in}$ the depairing rate, he obtains in the case of long junctions ($L>>L_{b}$):
\begin{equation}
G_{NS}=G_{NN}\frac{\int d\epsilon \frac{1}{4}A(\epsilon) f_{N,z}(\epsilon,V)}{2\sqrt{\int_{0}^{|V|} dV_{1}\int d\epsilon \frac{1}{4}A(\epsilon) f_{N,z}(\epsilon,V_{1})}}
\label{eq:Gns}
\end{equation}
with
\begin{displaymath}
\frac{1}{4}A(\epsilon)=Re \frac{\epsilon_{b} \gamma_{in}}{\epsilon^{2}+\gamma_{in}^{2}} \frac{\Delta^{2}}{\Delta^{2}-(\epsilon+i\Gamma_{s})^{2}} \theta(\Delta-|\epsilon|) 
\end{displaymath}
\begin{equation}
+Re\sqrt \frac{(\epsilon+i\Gamma_{s})^{2}}{(\epsilon+i\Gamma_{s})^{2}-\Delta^{2}} \theta(|\epsilon|-\Delta)
\label{eq:Gspect}
\end{equation}
\begin{equation}
\gamma_{in}=\frac{\hbar}{\tau_{in}}
\end{equation}
\begin{equation}
\epsilon_{b}=\frac{\hbar D}{\mathrm{L}_{b}^{2}} \textrm{ ,  } L_{b}=\sqrt{\frac{2 R_{b}}{R_{\Box}}}
\end{equation}
with $\Delta$ the superconducting gap, $\Gamma_{s}$ a small damping in the superconductor, $\tau_{in}$ the inelastic time and $L_{b}$ the barrier length, $R_{b}$ the tunnel barrier resistance and $R_{\Box}$ the sheet resistance of the normal part. $f_{N, z}=\frac{1}{2}[\tanh(\frac{\epsilon +eV}{2k_{B}T})-\tanh(\frac{\epsilon -eV}{2k_{B}T})]$ is the equilibrium distribution function.

The first term in $\frac{1}{4}A(\epsilon)$ (equation \ref{eq:Gspect}) describes the pair amplitude in the normal metal. It is non-zero (although small) in spite of the null value of the electron-electron interaction potential in this region: this is the proximity effect. As $\epsilon_{b} \rightarrow 0$ (the barrier is more and more opaque), the proximity effect reduces and the pair amplitude vanishes. In the same way, this pair amplitude is zero when there is no coherence effect ($\gamma_{in} \rightarrow +\infty$). The second term is the BCS normalized superconducting density of states, which leads to the usual conductance-voltage characteristics (tunnel Hamiltonian approach). At zero temperature and voltage, $G_{NS}=G_{NN}\sqrt{\epsilon_{b}/\gamma_{in}}$ as long as $\epsilon_{b} \textrm{, } \gamma_{in} << \Delta$. The width of the peak is roughly proportional to $\gamma_{in}$.

As we said in the introduction, S/Sm junctions are intermediate between good metallic junctions and oxyde tunnel barriers. We choose to treat our junctions within the SIN Volkov's model, taking into account that the absolute contact resistances are high and that estimation for the Schottky barrier gives poor interface transparencies. We will show that applying Volkov's model forces us to attribute a much larger sheet resistance for the silicon layer below the overlap  ($R_{\Box}^{*}$=315 $\Omega$) than for the bare silicon  ($R_{\Box}$=24 $\Omega$). The depairing rate below the interface is also found to be much lower than the phase breaking rate for the bare film. Within this theory, we stress that it is impossible to consistently describe the ZBA taking the silicon parameters in the bare film unchanged  below the overlap. This illustrates how the full quantitative analysis of the ZBA provides distinctive informations  about the microscopic parameters of the S/Sm contacts, otherwise unaccessible.

Figure \ref{fig:rt} shows the experimental data for the zero bias resistance versus temperature and the best Volkov's model adjustment. The parameter are the following: titanium nitride superconducting gap $\Delta$=0.21meV, damping factor $\Gamma_{S}/\Delta$=0.12, normal resistance $R_{NN}$=7.45$\Omega$, barrier transparency per channel $\Gamma=3.4\ 10^{-2}$, and finally the pair breaking energy $\gamma_{in}=60-85\mu eV$.

There are two separate sets of parameters in this list: $\Delta$, $\Gamma_{S}$, $R_{NN}$ are determined in the high temperature range (and also independently by the large voltage range in the differential resistance curves, see later on); $\gamma_{in}$ and  $\Gamma$ are fixed by the low temperature range of the curve. In principle, $\Gamma$ enters also as a parameter in the high temperature range, but due to the large value of  $\Gamma_{S}$, any relative small value of  $\Gamma$ satisfies the adjustement.

In the high temperature range (above 500mK), we  also fit the resistance versus temperature curves with the BTK model \cite{BTK}, using the same values for the parameters than for the Volkov's adjustment (see figure \ref{fig:rt}). Because in this range of energy the phase coherence is likely negligible, we find that both formalisms describe equally the experimental data. Below 500mK, coherence effects appear and the BTK model is no longer valid. 

We now comment on the various parameters: the damping factor $\Gamma_{S}$ is large probably because the disorder weakens the superconductivity at the interface. In particular, TEM image of the contacts indicates that a  TiSi$_{2}$ 40nm thick layer is produced by the thermal treatment at the interface \cite{Leti}. From the transparency, we estimate the barrier resistance $R_{b}=\frac{h}{2e^{2}} \frac{(\lambda_{F}/2)^{2}}{\Gamma}\simeq 18 \Omega .\mu m^{2}$. From the absolute value of $R_{NN}$, we can deduce the sheet resistance of the silicon {\it underneath} the interface:  $R_{\Box}^{*}=\frac{R^{2}_{NN} w^{2}}{R_{b}}$=315$\Omega$. This value is 15 times larger than the bulk value obtained with the TLM measurements ($R_{\Box}$=24 $\Omega$). Note that without the determination of $\Gamma$ by the Volkov's fit of the ZBA, another set of $\Gamma - R_{\Box}^{*}$ parameters could explain the absolute value of $R_{NN}$, namely for instance $R_{\Box}^{*} = R_{\Box} =24 \Omega $ and $\Gamma=\frac{h}{2e^{2}} \frac{(\lambda_{F}/2)^{2}}{R_{b}}=2.35\ 10^{-3}$ (with $\lambda_{F}=13.6nm$). This set will give a good BTK fit above $T=500 mK$, but the  $\Gamma$ is too small to give enough Andreev reflection to provide the ZBA. The ZBA is explained only if both the normal sheet resistance of the silicon layer under the overlap  and the transparency of the Schottky barrier are large enough. As said previously and also in reference \cite{Poirier97}, the semiconductor layer under the overlap could have a much larger sheet resistance than the native layer and is close to the metal/insulator transition ($k_{F} \ell_{e} \simeq 1$). In such a case, the phase breaking length should also strongly decrease in the vicinity of the overlap as compared to the bare film. 

Moreover, we obtain $\gamma_{in}=60-85\mu eV$, which leads to $\tau_{in} = \hbar / \gamma_{in}$=8-11ps. This value, which is an upper limit for $\tau_{\varphi}$, is smaller by two orders of magnitude as compared to $\tau_{\varphi}\simeq$2.7ns ($L_{\varphi}$=1$\mu$m at 240mK) estimated from magnetoresistance measurements in the bulk silicon layer. Interestingly, such small $\tau_{in}$ has been reported in tunnel experiments in copper wires \cite{Pothier94} or in gold wires from analysing the Josephson effect in SNS junctions \cite{Baselmans99}, and the discrepancy with phase breaking times measured by weak localization measurements also quoted. We do not know if it is coincidental, or if local measurements of $\tau_{in}$ near an interface generally lead to such underestimations as compared to bulk measurements. 

In summary, the precise analysis of the reflectionless tunneling effect leads us to deduce that the ZBA is not due to coherent backscattering over long distances (about $L_{\varphi}$=1$\mu$m) into the bulk silicon, but to coherent backscattering in a short  disordered layer.

Finally, we discuss the effect of an applied magnetic field  parallel or perpendicular to the junction. First, we measured the resistance-voltage characteristics at 30mK for various magnetic fields parallel to the interface (along $z$, see figure \ref{fig:tlm}). We note that the zero bias anomaly is divided by a factor two for an applied fields $H_{c}^{z} \simeq 200G$ and completely disappears at 400G.
We can estimate the field necessary to put one flux quantum $\Phi_{0}=h/e$ in a square of size $L_{in}^{2}$. Following Marmorkos et al \cite{Marmorkos93}, it gives the critical field to destroy the zero bias anomaly. Under the interface, we deduce from the fit of our experimental curves, $\tau_{in}$=8-11ps, so $L_{in}=\sqrt{D\tau_{in}}$=54-64nm with $D=3.67\ 10^{-4}m^{2}.s^{-1}$. Consequently, $H_{c}=\Phi_{0}/L_{in}^{2}\simeq$1T, which is much higher than the observed values of the field. Volkov et al. \cite{Volkov92} proposed another mechanism to destroy the coherence of the Andreev pairs. The magnetic field leads to a screening current, which gives a dependence of the phase of the order parameter with the coordinate $z$. This situation is comparable to Andreev interferometers \cite{Pothier94}: if the two electrons encounter a superconducting phase varying over $\pi$, destructive interferences occur and the ZBA disappears. Then, the depairing depends on the magnetic field through: 
\begin{equation}
\gamma_{in}(H)/\Delta = \gamma_{in}(0)/\Delta + \hbar D (\pi H \lambda/\Phi_{0})^{2}
\end{equation}
with $\lambda$ the London penetration depth of the superconductor. Using Volkov theory (see equation \ref{eq:Gns}), we calculate the value of the depairing rate at zero voltage for various fields and obtain the fitting curve: $\gamma_{in}(H)/\Delta=0.405 + 4.5\ 10^{-7} H(G)^{2}$ (see inset figure \ref{fig:r-v-h}), which leads to $\lambda=790nm$. This value seems rather high compared to $\lambda \approx 200nm$ for NbN. To calculate this penetration depth, we take the diffusion coefficient in the bulk silicon (see equation \ref{eq:D}) which maybe lower under the interface because of the disorder induced during annealing. Then, the penetration depth maybe overestimated. Nevertheless, this mechanism may cause the destruction of the interferences in our sample, since they are not sensitive to the decoherence induced by flux in electron-hole trajectories in the normal part (critical field too high).

But great care should be taken with the orientation of the magnetic field. We also measured the resistance-voltage characteristics at 30mK for various magnetic fields perpendicular to the interface (along $y$). The overall subgap resistance decreases for very small fields. Since titanium nitride is a type II superconductor, we attributed this effect to the appearance of vortices: the total resistance decreases because the junction normal resistance due to the vortices is less than the superconducting junction resistance. Secondly, the zero bias anomaly is divided by a factor two for a weak applied field $H_{c}^{y} \simeq 30G$ (see figure \ref{fig:r-v-h}). We can estimate the demagnetization factor under the interface. According to Zeldov et al. \cite{Zeldov94}, the field under the interface is given by $H_{interface}\simeq \sqrt{w/d_{TiN}} H_{applied} \simeq 10 H_{applied}$, with $d_{TiN}=100nm$ thickness of the TiN film and $w=1\mu m$. This value agrees with the ratio of characteristic field: $H_{c}^{y}/H_{c}^{z}\simeq 7$. Consequently, the ZBA for magnetic field applied perpendicularly to the interface disappears for weak field because of the demagnetization factor of the superconducting film.

The vortices may also explain the decreasing around 200G of the resistance background when the magnetic field is parallel to the interface: if the field is not strictly parallel to the interface, some vortices may appear with a small perpendicular component of the field.

\section{Electron heating and effective temperature}

\begin{figure}
\begin{center}
\psfig{figure=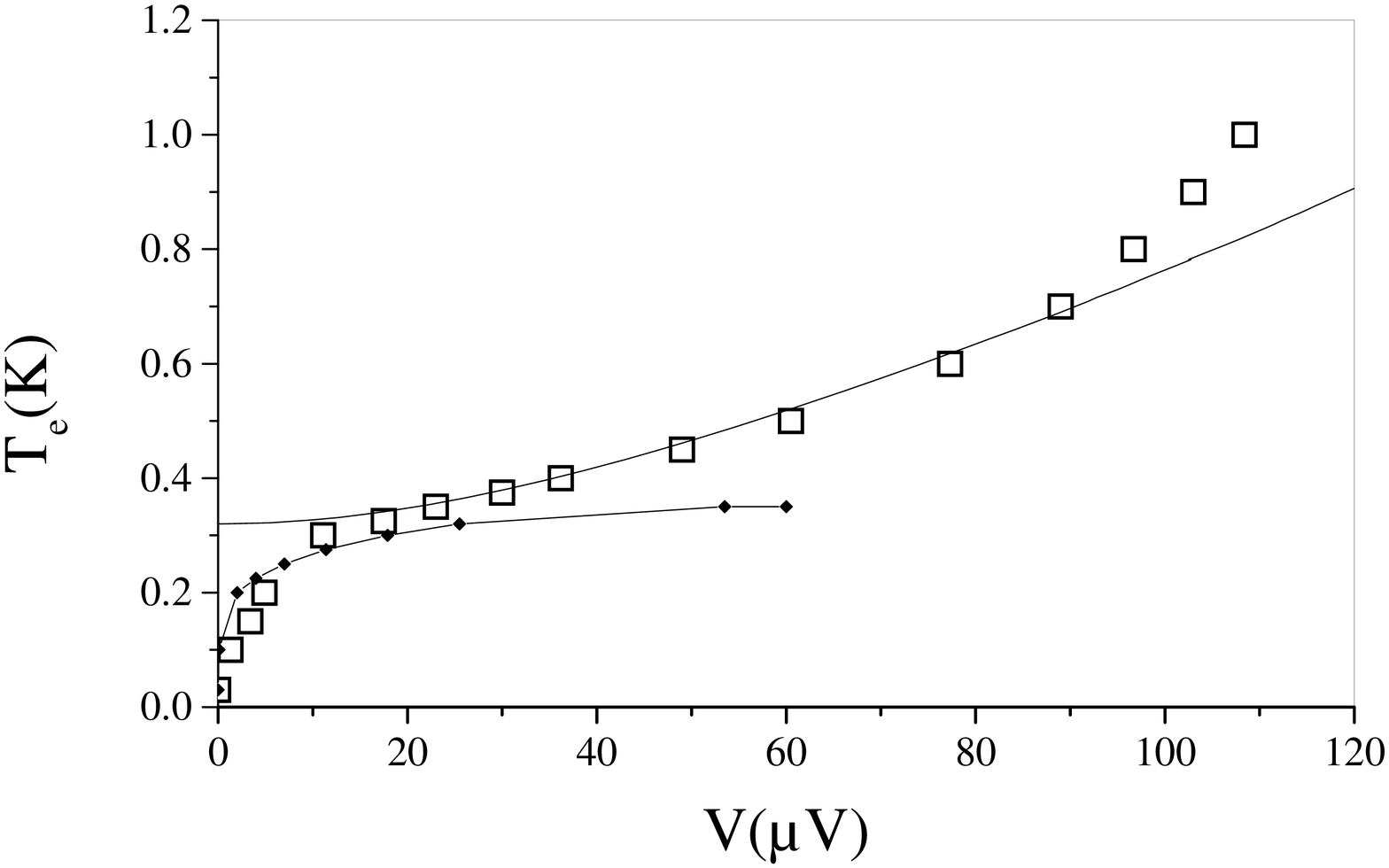,width=100mm}
\psfig{figure=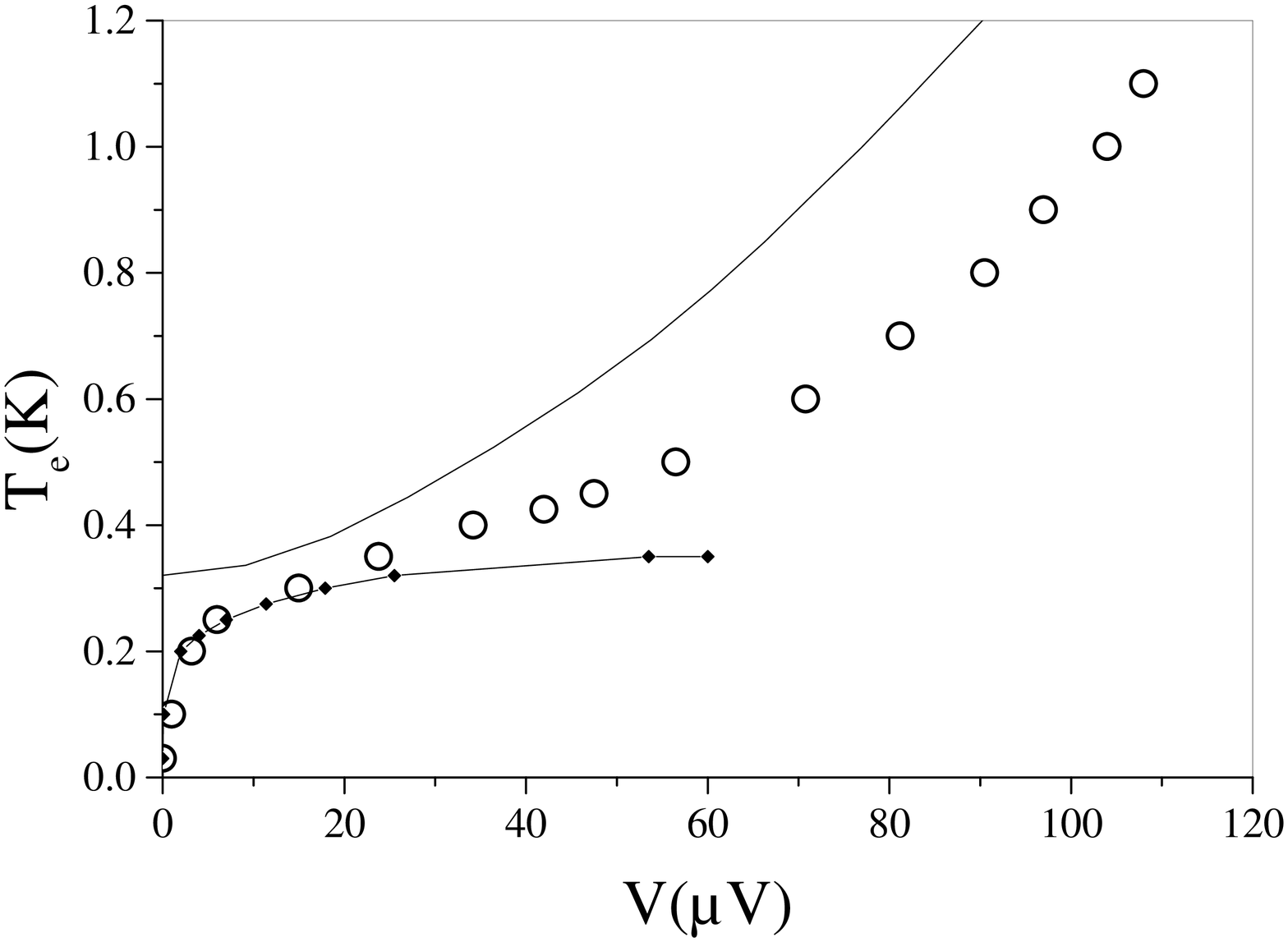,width=100mm}
\caption{Electronic temperature ($\Box$ points) deduced from comparison between theoretical (Volkov's model with previous parameters for R(V) curves; see text) and experimental R(V) curves of the 1$\times$10 $\mu m^{2}$ (top) and 20$\times$10 $\mu m^{2}$ (bottom) samples. The solid curve is the Wiedemann-Franz law $T_{e}(V)=\sqrt{T_{Si}^{2}+\frac{1}{4}\frac{3}{\pi^{2}}(\frac{eV}{k_{B}})^{2}}$ with $T_{Si}=320mK$. The diamond-curve is deduced from the dissipated power model, with $\Delta=0.21meV$, $\Gamma_{S}/\Delta=0.14$, $Z=5.3$, $R_{NN}=7.45\Omega$ and $T_{0}=30mK$.}
\label{fig:te-v}
\end{center}
\end{figure}

The Volkov's theory or the BTK theory are not able to fit our experimental resistance-voltage curves if we suppose an equilibrium Fermi distribution with a base temperature T$_{0}$. Paradoxically, we note that the maximum of differential resistance happens precisely at a voltage $V_{0}\simeq 20 \mu V$ such that $eV_{0} \simeq k_{B}T_{0}$, where $T_{0} \simeq 250mK$ is the temperature at which the zero bias resistance is maximum. Such coincidence happens also in the context of the re-entrance effect \cite{Courtois99a, Petrashov98} and has been observed in most of the reflectionless tunneling experiments \cite{Poirier97, Kastalsky91, Bakker94}. But this is not predicted in the theoretical models. From the model, we expect that the voltage brings to a much smaller suppression of the ZBA as compared to  experimental data (see figure \ref{fig:r-v}). A way to restore a good accordance between the model and the experimental voltage characteristics is to choose the effective temperature of carriers as an adjustable parameter \cite{Petrashov98}. At a fixed voltage, we find the elevated temperature such that the model gives precisely the measured resistance value. Therefore, we construct the variation of the effective temperature $T_{eff}$ as a function of the applied voltage (figure \ref{fig:te-v}). We attribute this temperature to the carriers in the doped silicon film in between the two TiN electrodes.

What is happening first at very low temperature when a finite voltage is applied to the TiN/Si n$^{++}$/TiN sample? Some Joule power is dissipated in the silicon part which is very hardly evacuated either in the phonon bath or in the contacts because the electron-phonon coupling is very small at low temperature and the Andreev thermal resistance at each N-S interface is very large \cite{Andreev64}. Consequently the electrons inside the silicon are overheated. Overheating has two causes. First, the Andreev thermal resistance, which depends exponentially on the effective temperature, is mainly responsible for the elevated temperature at low voltage, and one can neglect the gradient of temperature in the silicon between the superconducting contacts. Secondly, at high voltage, this gradient is non negligible and given by the Wiedemann-Franz law. We suppose  that the inelastic electron-electron scattering time is short as compared to the length of the sample ($1-20 \mu m$), to have an quasi-equilibrium Fermi distribution with an effective temperature.

At low voltage, below 20 $\mu$V, the electronic temperature increases very rapidly up to 320mK.  Hoss et al. \cite{Hoss99} observed a very similar behavior in  $1 \mu m$ long Nb/Au/Nb samples at low currents.  To understand this results, we used a BTK-based model of dissipated power \cite{Hoss99}. The power dissipated through a NS interface is given by:
\begin{equation}
P(T_{e},V_{i})=\frac{2\,G_{NN}}{e^{2}} \int_{-\infty}^{+\infty} d\epsilon \ \epsilon \, [f(\frac{\epsilon-eV_{i}}{k_{B}T_{e}})-f(\frac{\epsilon}{k_{B}T_{0}})][1-A(\epsilon)-B(\epsilon)]
\end{equation}
with $T_{e}$ and $T_{0}$ electronic and phonon temperatures, $V_{i}$ voltage drop at the interface, A and B the Andreev and normal reflexion probabilities \cite{BTK}, $G_{NN}=1/R_{NN}$ the normal conductance. We fit the curve with the parameters used earlier in the BTK fit: $\Delta=0.21meV$, $\Gamma_{S}/\Delta=0.14$, $Z=5.3$ (which gives a barrier transparency of $3.4\ 10^{-2}$), $R_{NN}=7.45\Omega$ and $T_{0}=30mK$. We assume that all the electric power is dissipated through the two NS interfaces, i.e. that the electron-phonon length at this temperature exceeds the distance between the two superconducting interfaces. Then, by equating $P(T_{e},V_{i})=V_{total}I$, with $V_{total}$ the total voltage and $I$ the current across the sample, we obtain $T_{el}(V_{i})$.

These parameters describe the experimental points of figure \ref{fig:te-v} at low voltages. Since the thermal resistance of the N/S interface decreases exponentially with temperature, for larger effective temperatures the heat is rapidly evacuated in the superconducting electrodes and a thermal equilibrium is reached, giving the saturation of the effective electronic temperature supposed constant in the silicon to around 300mK (solid symbols in figure \ref{fig:te-v}).

At higher voltages, the effective temperature continues to increase. Our hypothesis of a constant effective temperature in the silicon should fail. The simplest analysis is to consider the Wiedemann-Franz law, globally for the whole TiN/Si n$^{++}$/TiN sample. In the 1$\times$10 $\mu m^{2}$ sample, it follows a Wiedemann-Franz law (interacting hot electron regime) $T_{e}(V)=\sqrt{T_{Si}^{2}+\frac{1}{4}\frac{3}{\pi^{2}}(\frac{eV}{k_{B}})^{2}}$, with $T_{Si}=320mK$ temperature of the electron at the interface and V the total voltage applied to the SNS system (see figure \ref{fig:te-v}). In the 20$\times$10 $\mu m^{2}$, the effective temperature is less than predicted by the Wiedemann-Franz temperature. Heslinga et al \cite{Heslinga92} give an approximation of $L_{e-ph} \simeq 2 \mu m$ at 320mK. This value is intermediate between the length of our two samples and explains their different behaviors. In the 20$\times$10 $\mu m^{2}$ sample, as the length of the sample matches the electron-phonon length, electrons can interact with phonons and be cooled under the Wiedemann-Franz temperature, whereas in the 1$\times$10 $\mu m^{2}$, electrons are only heated by electron-electron interaction and the Wiedemann-Franz law is valid.

Finally, we test our estimation of the effective temperature using  the BTK model. On figure \ref{fig:r-v} (bottom), we observe a much better accordance with the data if we introduce the effective temperature given by the  Wiedemann-Franz law than if we use the base phonon temperature. For consistency, all the parameters except the temperature are obtained from the BTK fit of $R(V=0)$ versus  temperature.

\section{Conclusion}

In conclusion, we measured the differential resistance versus temperature, applied voltage and magnetic field of semiconductor/superconductor junctions. TiN/Si n$^{++}$ heterostructures proved to be a new and interesting tool for the study of proximity effect. We observed a zero-bias anomaly due to reflectionless tunneling. By comparing our results to a proximity effect theory \cite{Volkov93}, we find that the sheet resistance of the silicon underneath the interface is much larger than the bulk value and that the quasiparticule lifetime is much shorter than in bulk silicon. We explain these discrepancies by disorder induced by annealing during the process. Interestingly, the quasiparticule lifetime is comparable to the electron-electron interaction time deduced from tunnel experiments in copper wires \cite{Pothier94} or in gold wires from analysing the Josephson effect in SNS junctions \cite{Baselmans99}. 
At very low temperature and finite voltage, the effective electronic temperature in the silicon is much higher than the phonon temperature. This is valid for separation between superconducting contacts as large as $20 \mu m$. The effective temperature increases very rapidly at low voltage and then follows a Wiedemann-Franz law. This behavior is well explained by a model of heat dissipation through a N/S interface.

This rapid raise of the temperature of the carriers with the injected power is one of the interesting properties of this system. It could be used to make a bolometer, by measuring directly the zero bias conductance as function of the absorbed power in the silicon \cite{Nahum93}.

\section{Aknowledgments}

We would like to thank MM. Deleonibus and Demolliens (CEA/Leti) for providing the TiN/Si n$^{++}$ bilayers and the use of the PLATO facilities.

\end{document}